\documentclass[11pt,reqno]{amsart}
\usepackage[utf8]{inputenc}
\usepackage{tikz-cd}
\usepackage{pdfpages}
\usepackage{graphicx}
\usepackage{amssymb,amsmath, amscd, latexsym,amsfonts,bbm, amsthm,stmaryrd, enumerate,phaistos}
\usepackage{yhmath}
\usepackage{comment}\usepackage{todonotes}\usepackage[all]{xy}
\usepackage[vcentermath]{youngtab}

\usepackage{amsmath, amssymb, latexsym, enumerate,  graphicx,tikz, stmaryrd,pifont,ifsym}
\usetikzlibrary{arrows,decorations,decorations.pathmorphing,positioning}
\usepackage{mdframed}
\usepackage{mdwlist}
\usepackage[pdftex]{hyperref}
\usepackage{amscd,mathrsfs,epic,empheq,float}
\usepackage{bbm}
\usepackage{cases}
\usepackage[all]{xy}

\usepackage{tikz}
\usepackage{tikz-cd}

\theoremstyle{definition}

\usepackage{pict2e}
\def\StrangeCross

\usepackage{amsfonts}
\usepackage{amssymb}
\usepackage{amscd}
\usepackage{amsmath}
\def\bea{\begin{eqnarray}}
\def\eea{\end{eqnarray}}
\def\nn{\nonumber}

\usepackage{pict2e}

\newtheorem{theorem}{Theorem}[section]

\newtheorem{lemma}[theorem]{Lemma}
\newtheorem{proposition}[theorem]{Proposition}
\newenvironment{exafont}{\begin{bf}}{\end{bf}}

\newtheorem{corollary}[theorem]{Corollary}
\theoremstyle{definition}
\newtheorem{definition}[theorem]{Definition}
\newtheorem{remark}[theorem]{Remark}
\newtheorem{example}[theorem]{Example}

\begin{document}

\title{Electrical varieties as vertex integrable statistical models 
}
\author{Vassily Gorbounov and Dmitry Talalaev}
\address{V. G.: King's College, University of Aberdeen, AB24 3FX, UK,  HSE University, Usacheva 6, Moscow and \
Moscow Institute of Physics and Technology, Laboratory of algebraic geometry and homological algebra, Dolgoprudny, Russia. }
\email{vgorb10@gmail.com}

\address{D. T.: 119991, Moscow State University, Russia, Vorobievy Gory; Institute for Theoretic and Experimental Physics, Russia, Moscow and Demidov State University, Russia, Yaroslavl} \email{dtalalaev@yandex.ru}

\maketitle
\tableofcontents
\bigskip
\begin{abstract}
We propose a new  approach to studying electrical networks interpreting the Ohm law as the operator which solves certain Local Yang-Baxter equation. Using this operator and the medial graph of the electrical network we define 
a vertex integrable statistical model and its boundary partition function. This gives an equivalent description of 
electrical networks We show that in the important case of an electrical network on the standard graph introduced in \cite{CIM}, the response matrix of an electrical network, its most important feature, and the boundary partition function of our statistical model can be recovered from each other.

Defining the electrical varieties in the usual way we compare them to the theory of the Lusztig varieties developed in \cite{BFZ}.
 In our picture the former turns out to be a deformation of the latter. We describe how  our approach produces
new interesting mathematical structures on the  electrical varieties.  Our results should be compared to the  earlier work started in \cite{LP} on the connection between the Lusztig varieties and the electrical varieties.
\end{abstract}
\section{Introduction}
The theory of cluster algebras \cite{FZ}, \cite{FG}, \cite{GSV} and the theory of directed networks, not necessarily planar, developed in \cite{Post} grew out of a very interesting paper \cite{BFZ} where the authors studied the Lusztig variety $\frak  L$, 
the variety of parametrisations of the unipotent group $U_n$. The following features of this variety are important for us: 
\begin{itemize}
\item the set of toric charts labelled by reduced words for the longest permutation coming out of the factorization of a unipotent matrix into a product of the Jacobi matrices
\item the set of directed graphs labelled by the same set as the charts, these directed graphs are obtained out of the pseudo-line arrangements defined by reduced words for the longest permutation
\item  a supply of functions defined as sums of weights of directed paths in these graphs which can be packed into a unipotent upper triangular matrix
\item the transition maps between the charts correspond to the local  ``Yang-Baxter'' mutations
and a particular solution to the tetrahedron Zamolodchikov equation defines the formulas for these transition maps 
\item  the nil Temperley-Lieb algebra related to this solution
\item a particular set of functions $M_L$ labelled by the subsets of the set $[1,n]$  which generate the ring of functions of the Lusztig variety and obey a Pl\"ucker type relation
\item the natural action of the symmetric group on the electrical varieties associated with important classes of graphs analogous to the discrete Toda lattice
\end{itemize} 
These led among other things to the discovery of the concept of the cluster algebra and 
the theory of directed networks.

Let us look at a different problem. Suppose we are given a connected graph $\Gamma$, a part of its vertices is called the boundary the rest are forming the interior and a function $\gamma$ from the edges of $\Gamma$ to the non-zero complex numbers. If $\gamma$ takes only positive real values the pair $(\Gamma,\gamma)$ defines an electrical network with the conductivity $\gamma$ on $\Gamma$ \cite{CIM}, \cite{CdV}. 
Apply the electrical potential  to the boundary vertices then the Ohm law and the Kirchhoff law imply that the current through the internal vertices is zero and therefore we obtain a linear map sending the values of the potential on the boundary to the currents from the boundary to the network. The matrix of this linear map is called the response matrix of the network. One naturally obtains a {\it variety} $\frak T$  of electrical networks in this situation. Indeed the choice of the conductivity function, not necessary positive, should be treated as a point in the toric chart labelled by $\Gamma$. The graph of the network can be changed locally using the so called ``star-triangle'' transformation which involves three edges incident to each other in two different ways. If one changes the value of the conductivity function on these edges appropriately, obtaining a  new network,  the responses matrix of this network is equal to the response matrix of the old network. This change of the conductivity defines the transition map between the toric charts. In what follows we use the fact that this transformation produces a solution for the Zamolodchikov tetrahedron equation. The response matrix $M_R$ is therefore a matrix-valued function on our variety. Studying the variety $\frak T$, which we will call the {\it electrical variety}, is the natural task.  Moreover it is evident from the description of the electrical varieties  that one should search for the connection between them and  the Lusztig varieties. 

The first work in this direction was done in  \cite{LP} where the authors clearly introduced the idea that the theory of the electrical varieties is a deformation of the theory of the Lusztig varieties. Using a solution of the Yang-Baxter type equation the authors introduced the electrical Lie group which acts on the set of electrical varieties and is closely related to the symplectic group. This idea was further  developed in the subsequent work \cite{LP4}, \cite{LP5}, \cite{Lam} leading to a number of significant developments in the theory of the electrical varieties.

We suggest a new approach to studying the electrical varieties  by introducing a new complete electrical invariant, the data equivalent to the response matrix. Consider the medial graph of $\Gamma$. The ``star-triangle'' transformation of the original graph corresponds to the ``Yang-Baxter'' mutation in the medial graph. It turns out that  the Ohm law defines an operator which solves the Local Yang-Baxter equation with spectral parameters related
by the Zamolodchikov map - a solution for the Zamolodchikov set-theoretical equation. This map was known for a long time, it has been used earlier in connection to $(2+1)$ integrability in \cite{KKS} and it is different from the one used in \cite{LP}.
We will call our operator the Ohm-Yang-Baxter operator and denote it by $\phi$. 
Using the Ohm-Yang-Baxter operator and the medial graph of $\Gamma$ we will define a {\it vertex integrable statistical model} and a matrix-valued function $M_B$ on $\frak T$ which is the {\it boundary partition function}  of 
our vertex statistical model. 

In the paper we will show that this statistical model gives an equivalent description of the electric variety at least for some important class of electrical networks introduced in \cite{CIM}. This class of networks is special because the inverse problem for these networks can be solved, namely, given the response matrix one can recover the conductivity function of the network. The graph of such a network is called the standard graph in \cite{CIM}. 
We will give precise formulas relating the matrices $M_R$ and $M_B$ and solve the inverse problem for $M_B$ giving therefore a new solution for the inverse problem of the electrical networks on the standard graphs.

The medial graph of the standard graph turns out to be exactly the pseudo-line arrangement for a certain reduced word for the longest permutation used in \cite{BFZ}. 
It allows us to compare our approach to the theory of the electrical varieties for these networks to the theory of the Lusztig varieties discovering all the features we listed above with the appropriate modification.

\begin{itemize}
\item the solution of the Local Yang-Baxter equation coming out of the Ohm law is in fact the deformation of the  Lusztig solution
\item the matrices $M_B$ belong to the symplectic group and are the deformation of the unipotent upper triangular matrices
\item  for $n$ even the partition function $M_B$ gives an embedding of the variety $\frak T$ to the symplectic group $Sp(n)$
\item the Temperley-Lieb algebra $TL(0)$ at a root of unity of which the nil Temperley-Lieb algebra is a degeneration 
\item the ``Pl\"ucker coordinates'' will be constructed making the algebra of functions of the variety $\frak T$ a cluster like algebra which is the deformation of the cluster algebra of functions on the Lusztig variety
\item  the discrete Toda system analog naturally acts on the electrical variety defined by the rectangular lattice and the standard graphs.
\end{itemize}

Concluding the introduction we will point out two things.
Firstly,  almost all of the  items on the list above also appeared in the work \cite{LP}, \cite{LP4}, \cite{LP5}. However the connection of them to our work needs to be clarified. Secondly, a different statistical model approach to the electrical varieties via the dimer models was developed  and the cluster algebra structure on the electrical varieties was introduced in \cite{GK} and \cite{KW}. Our statistical  model is an integrable model of the vertex type and we use its specific properties to derive the features listed above.
Combining all these points of view and establishing the precise relationship between them should be an interesting task, we hope to return to it later.

The paper is organized as follows: the sections $1$ and $2$ serve to introduce the definition of the electrical variety and the notion of the vertex statistical model. The major results of the paper are situated in the sections $3$ and $4$. Part of the material in the sections $5$ and $6$ will be known to  experts. We included it as an illustration of our point of view on the idea that the electrical varieties are the deformation of the Lusztig varieties. 

\subsection*{Acknowledgements}
The authors are grateful to Arkady Berenstein, Azat Gainutdinov, Michael Gekhtman, Gleb Koshevoy and Vladimir Roubtsov for generously sharing knowledge and ideas and to Thomas Lam and Pavlo Pylavskyy who read the first draft of the paper and made a number of critical but helpful remarks and comments.

The work of V.G. was carried out within the HSE University Basic Research Program
and funded (jointly) by the Russian Academic Excellence Project `5-100'. 

The
work of D.T. was carried out within the framework of the State Program of the Ministry
of Education and Science of the Russian Federation, project No. 1.13560.2019/13.1, and was also
partially supported by the RFBR grant 17-01-00366 A.  D.T. is grateful for the Th\'el\`eme atmosphere of the IHES where the part of this work was done.

\section{The definition of the electrical variety}
The definition of electrical variety has appeared already in \cite{GK}, \cite{Lam}, and \cite{LP}, so we just recall it here. 

Suppose we are given a  pair $(\Gamma,\gamma)$, where $\Gamma$ is a connected graph, a subset of its vertices is labelled as the boundary vertices or nodes, the rest are forming the set of the interior vertices and  $\gamma$ is a function from the set $E(\Gamma)$
 of edges of $\Gamma$ to real numbers. It defines an electrical network with the conductivity function $\gamma$ on $\Gamma$ if $\gamma$ takes positive values. 

Let us denote by $\frak N(\Gamma)$ the set of graphs which can be obtained out of $\Gamma$   by applying the local star-triangular transformation or its inverse which are  pictured below:
\begin{figure}[h!]
\center
\includegraphics[width=60mm]{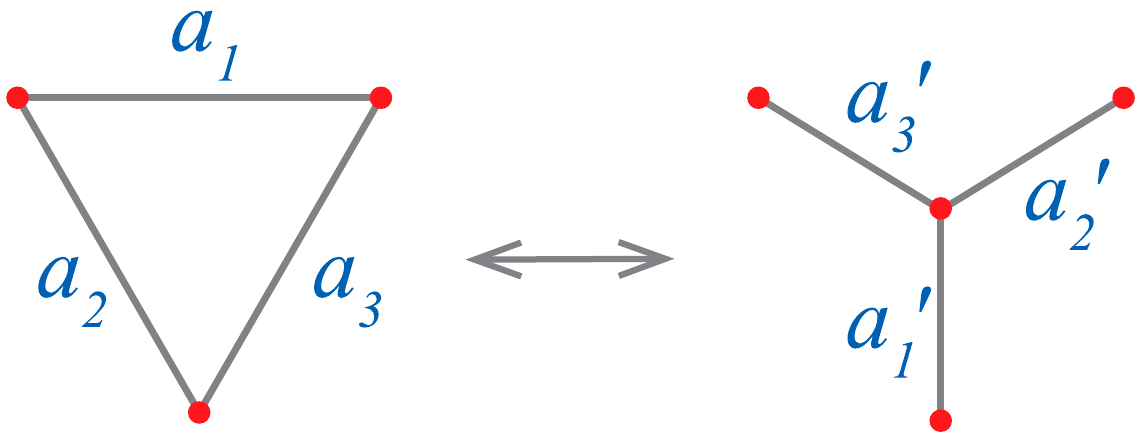}
\caption{Star-triangular transformation}
\label{fig:st-tr}
\end{figure}

Note that these transformations change the number of interior vertices by one. 

For a graph $\Gamma'$ obtained from $\Gamma$ by such a transformation  and a function $\gamma$ on the set of edges of $\Gamma$ define a function $\gamma'$ on the edges of $\Gamma'$ as follows: 
the value of $\gamma'$ on all the edges is the same as the value of 
$\gamma$ except for the new edges $a'_1,a'_2,a'_3$ which replace the edges $a_1, a_2, a_3$ where
\[\gamma(a_i)\gamma(a'_i)=\gamma(a_1)\gamma(a_2)+\gamma(a_1)\gamma(a_3)+\gamma(a_2)\gamma(a_3)=\frac
{\gamma(a'_1)\gamma(a'_2)\gamma(a'_3)}{\gamma(a'_1)+\gamma(a'_2)+\gamma(a'_3)}\]
$i=1,2,3$.
\begin{definition}We will say that  $(\Gamma',\gamma')$ and $(\Gamma,\gamma)$ are related by the {\it  star-triangular mutation}.
\end{definition}
Let $|E(\Gamma)|=m$.
\begin{definition} 
The electrical variety $\frak T$ associated with the pair $(\Gamma$ is a collection of charts $\{{\bf t}_k \cong (\Bbb C^*)^m\}$ labelled by the elements of the set $\frak N(\Gamma)$. Two points $\gamma$ and $\gamma'$ in the charts labelled by $\Gamma$ and $\Gamma'$ are glued together by the above maps if the pairs $(\Gamma,\gamma)$ and $(\Gamma',\gamma)'$ are related by the star-triangular mutation.
\end{definition}
\begin{definition}
 Suppose $(\Gamma,\gamma)$ is a connected electrical network, with $n$ boundary vertices numbered $v_1,...,v_n$. The Kirchhoff matrix $K = K(\Gamma,\gamma)$ is the $n\times n$ matrix constructed as follows. 
 \begin{itemize}
 \item If $i \not= j$ then $K_{ij}=-\sum \gamma(e)$, where the sum is taken over all edges $e$ joining $v_i$ and $v_j$. If there is no edge joining $v_i$ to $v_j$, then $K_{ij}=0$
 \item  $K_{ii} = \sum\gamma(e)$, where the sum is taken over all edges $e$ with one endpoint at $v_i$  
 and the other endpoint not $v_i$. 
 \end{itemize}
 \end{definition}
 
The Schur complement $M_R(\Gamma,\gamma)$ in the Kirchhoff matrix associated with pair $(\Gamma,\gamma)$ to the submatrix of the interior vertices of $\Gamma$ is called the {\it response matrix} \cite{CIM}.  Its role in the theory of electrical networks is described in the introduction. If it does not lead to confusion we will denote the response matrix simply by $M_R$. Note that due to the properties of the Schur complement the matrix $M_R$ is singular because the Kirchhoff matrix is singular.

The response matrix has many remarkable properties. The following will be especially important for us
\begin{proposition}
The matrix $M_R$ does not change under star-triangular mutations of the pair $(\Gamma,\gamma)$ and therefore it defines a function on $\frak T$ with values in matrices.
\end{proposition}

\section{Electrical varieties as vertex statistical models}
\subsection{Vertex statistical models}
\label{dir-net}
The definition below is the usual definition from statistical mechanics adapted to our situation.
\begin{definition}
By a vertex statistical model, the vertex model for short,  we mean a pair  $(N,{\bf X})$, where  $N$ is a finite directed graph without directed cycles and multiple edges situated  in the disk whose set of
vertices $V$ is divided into two subsets, the  interior vertices $V_0$ and the boundary vertices $V_1$, and ${\bf X}$ is a collection of  $2\times 2$  matrix $X_a$ with the coefficients in some ring, one for each interior vertex $a\in V_0$. 
The boundary and the interior vertices are situated on the boundary and in the interior of the disk respectively. The boundary vertices have the degree one and the interior vertices have degree 4. We assume also 
that each interior vertex to have in-degree $2$ and out-degree $2$, and the incoming edges and the outgoing edges are adjacent, as shown in the figure \ref{fig-cross}. 
\begin{figure}[h!]
\center
\includegraphics[width=25mm]{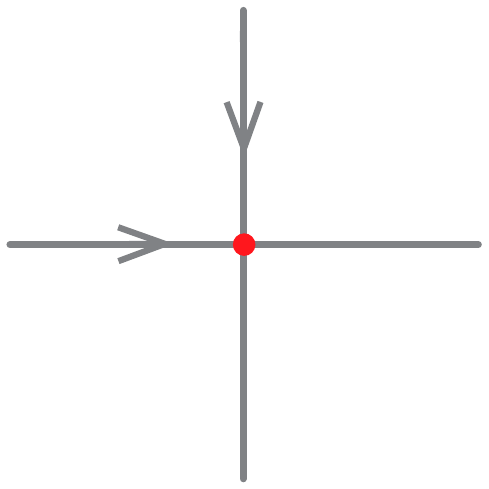}
\caption{Cross}
\label{fig-cross}
\end{figure}
\end{definition}

For each cross as above there are four ways a path can traverse  through the vertex:  $NS$, $NE$, and $WS$, $WE$, then  the rows of the matrix $X_a$ can be viewed as the weights of these paths through it.

Since the graph is directed the boundary vertices are naturally divided into the sources and the sinks.
\begin{definition} The {\it boundary partition function} $M_B$ for such a vertex model $(N,{\bf X})$ is a matrix which  rows are labelled by the sources, the columns are labelled by the sinks.
Given a source $i$  and  a sink $j$ define $(M_B)_{ij}$ as the sum of the weights of all the paths between $i$ and $j$. 
\end{definition}\label{partition function}

We will offer now a different way of calculating the matrix $M_B$ which shows that this matrix is indeed a boundary partition function of a vertex statistical model. Let us call {\em a strand} a path which goes through the vertex from the North to the South or from the West to the East. Let us enumerate all the strands of the graph $N$. Then to each vertex $a$ which is the intersection of say the  $i$-th and the $j$-th strands we assign a block matrix $(X_a)^{ij}$ which is the identity matrix except the submatrix with $i$-th and $j$-th rows and columns which is $X_a.$

Since the graph $N$ does not have directed cycles there is a natural partial order on the vertices: $a>b$ if there is an oriented edge connecting $b$ to $a$. Consider the product of the matrices $(X_a)^{ij}$ (over the ordered  set of vertices) in which the order of the factors agrees with the order of the vertices who label the factors
\bea
M=\prod_{a\in V}(X_a)^{ij}.\nn
\eea  
\begin{lemma}
The matrix $M$ is well defined and moreover $M=M_B$.
\label{lem_prod}
\end{lemma}
\begin{proof}
Although this statement is well known we propose here our interpretation a proof of it. 

First, note that from the construction of the matrices $(X_a)^{ij}$ it is clear that if two vertices $a$ and $b$ are not comparable in the partial order then the matrices $(X_a)^{ij}$ and $(X_b)^{kl}$ commute, hence the matrix $M$ is well defined.

Now consider the sum of the weights of the paths going from a source vertex to a sink vertex. The matrix elements of $(X_a)^{ij}$ define the weights of all  the ways a path can travel along the strands intersecting at $a$. 
At each point there are four such ways as the figure \ref{fig-cross} shows and we will identify them with the matrix elements of $(X_a)^{ij}$ as follows before
\bea\label{lus}
\varphi_L=\left(
\begin{array}{cc}
WE& WS\\
 NE &NS
\end{array}
\right).
\eea
It is clear now that to each such path we can find a unique summand in the expression of the appropriate matrix element of $M$. 
\end{proof}

We can perform the following local transformation on our vertex model. It changes the graph $N$ as follows:
\begin{figure}[h!]
\center
\includegraphics[width=85mm]{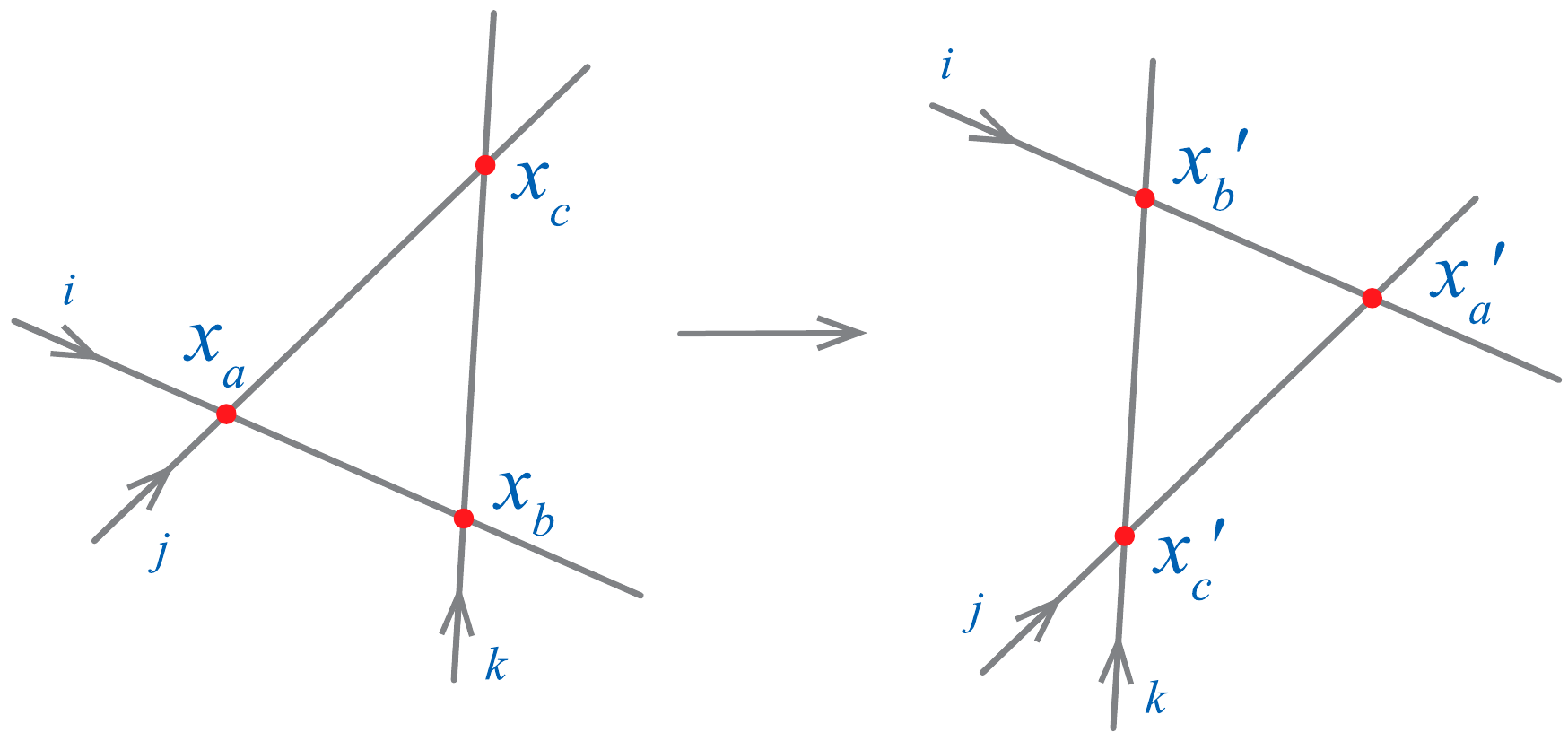}
\caption{Yang-Baxter mutation}
\label{fig:YB}
\end{figure}

In words it moves the line $k$ through the intersection point of the lines $i$ and $j$. Here we denote a vertex by the labels of the lines which intersect at this vertex. 

Suppose also that we find three matrices $ (X_{a'}), (X_{b'}), (X_{c'})$ such that the following {\it Local Yang-Baxter} equation holds:
\[(X_{a})^{ij}(X_{b})^{ik}(X_{c})^{jk} = (X_{c'})^{jk}(X_{b'})^{ik}(X_{a'})^{ij}.\] 
The upper indices as usual show the pairs of indices in $\Bbb C^n$ the appropriate operator acts on. Our transformation replaces the matrices  $X_{a}, X_{b}, X_{c}$ in the set $\bf X$ by the primed matrices
$ X_{a'}, X_{b'}, X_{c'}$. The rest of the network data remains unchanged. 
\begin{definition}
Such a transformation of the vertex model $(N,{\bf X})$ we will call the Yang-Baxter mutation. 
\end{definition}
Denote by $\frak N(N,{\bf X})$ the set of all vertex models obtained from $(N,{\bf X})$ by the Yang-Baxter mutations. 

\begin{proposition} The Yang-Baxter mutation does not change the boundary  partition function $M_B$.
\end{proposition}
\begin{proof}
 This is obvious due to lemma \ref{lem_prod}. Indeed, the products of the local vertex matrices for both vertex models coincide as a consequence  of the local Yang-Baxter equation. 
\end{proof}

\subsection{Directed graph associated to electrical network}
We will show now how to associate a vertex model in the above sense to an electrical network. 
First we will define a directed graph (naturally) related to an electrical network.

Suppose $\Gamma$ is a planar graph with no loops, with $n$ nodes; $\Gamma$ is embedded in the plane so that the nodes $v_1, v_2, . . . , v_n$ are in clockwise 
order around a circle $C$ and the rest of 
$\Gamma$ is in the disc $D$ bounded by $C$.

Recall the construction of the {\it medial graph} $\Gamma^M$  of  $\Gamma$. It depends on the embedding of $\Gamma$  in the disc. 
For each edge $e$ of  $\Gamma$ let $m_e$ be its midpoint. Place $2n$ points $t_1, t_2, . . . t_{2n}$ on $C$ so that:
\[
t_{1}< v_1< t_2< t_2< v_2< t_3<. . . < t_{2n-1}< v_n< t_{2n}
\]

The vertices of  $\Gamma^M$ are the points $m_e$ for all edges $e$ in  $\Gamma$ and the points $t_i$ for $i= 1,2, . . . ,2n$.

\begin{figure}[h!]
\center
\includegraphics[width=60mm]{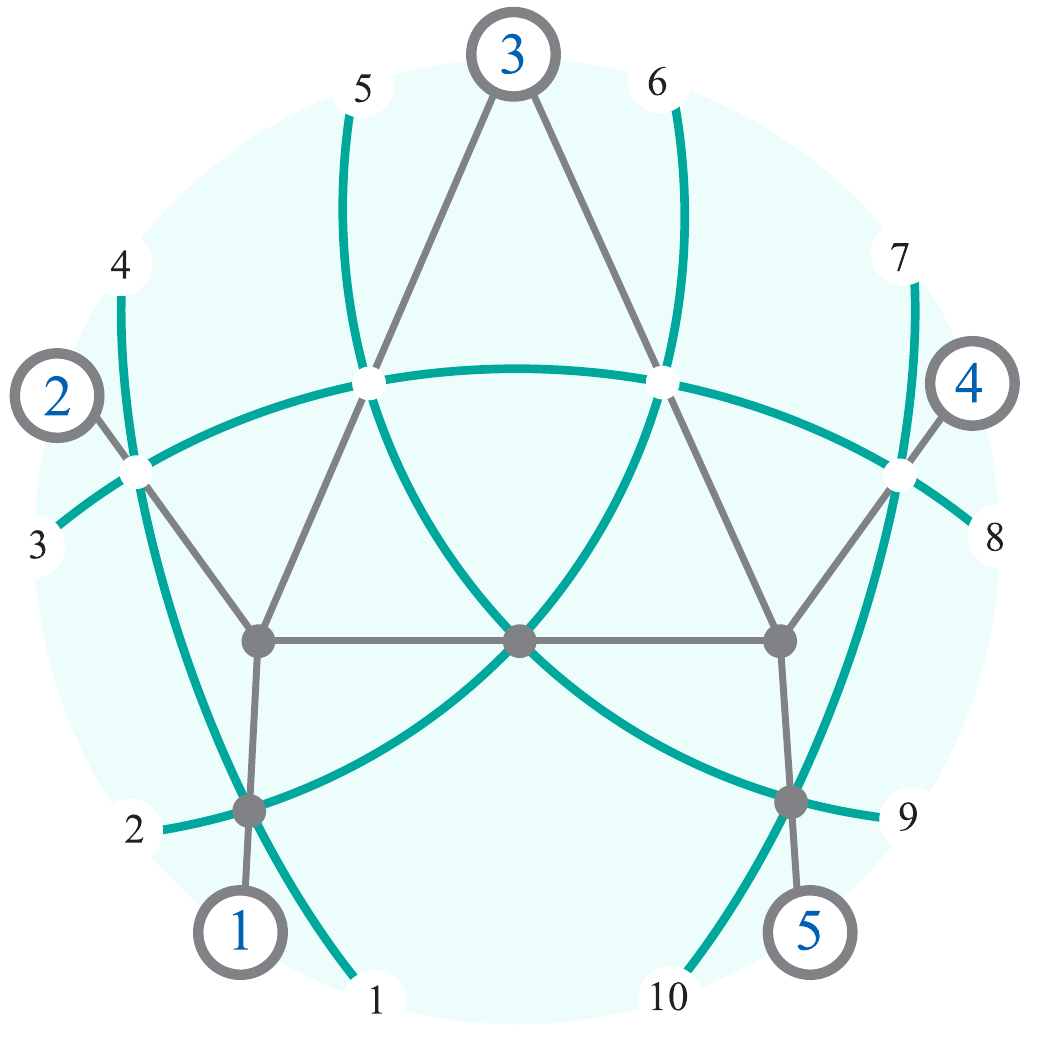}
\caption{Network}
\label{fig-network1}
\end{figure}

If $e$ and $f$ are edges in  $\Gamma$ with a common vertex, and which are incident to the same face, the interval joining $m_e$ and $m_f$ will be an edge in  $\Gamma^M$. For each point
$t_j$ on the boundary circle, there is one edge as follows: the point $t_{2i}$ is joined by an edge to $m_e$ where $e$ is the edge whose endpoint is  $v_i$ which comes first after the line $v_i t_{2i}$ in clockwise order around
$v_i$; the point $ t_{2i-1}$ is joined by an edge to $m_f$ where $f$ is the edge whose endpoint is  $v_i$ which comes first after the line $v_i t_{2i-1}$ in counter-clockwise order around $v_i$ as shown on the picture
\ref{fig-network1}.
  
As it is true for any medial graph the interior vertices of $\Gamma^M$ are of degree $4$.
The {\it strand}, as in section \ref{dir-net}, is a path in $\Gamma^M$ which connects two nodes of $\Gamma^M$ and goes straight through any interior vertex. This defines a perfect matching on the set of nodes of $\Gamma^M$ and an orientation on any strand as well. Indeed each strand has the endpoints labelled by the unique pair of distinct numbers between $1$ and $2n$. We will think of this pair as defining the direction going from the smaller to the larger label.
Thus the medial graph has the orientation defined by the order on the set of nodes and the nodes. 

The conductivity function $\gamma$ on the edges of $\Gamma$ translates into a function on the vertices of $\Gamma^M$, which we denote by the same letter $\gamma$.

The graph $\Gamma^M$ has an important additional structure: its vertices are naturally coloured by one of two colours, the black or the white.
If the vertex belongs to an edge of $\Gamma$ then the line this edge belongs to breaks the strands which intersect in the vertex in two possible ways: either the sources of the strands sit in different half planes bounded by
 that line or in the same.  In the first case we colour the vertex black and in the second white, see figure \ref{fig-network1}. 

\subsection{The Ohm-Yang-Baxter operator}
 The next step in the construction of the vertex model is to define the set of matrices $\bf X$ labelled by the vertices of the graph $M^{\Gamma}$ which solve the Local Yang-Baxter equation. 
This will be done using the major feature of electrical networks, the Ohm law.  

The graph $\Gamma^M$ is a directed graph whose internal vertices are identified with the edges of $\Gamma$. The conductivity function $\gamma$ therefore is a function on the set of the vertices of $\Gamma^M$. 

Recall that the vertices of $\Gamma^M$ are coloured either by the black or by the white colour. We will describe the two types of variables one can attach to the vertices of 
$\Gamma^M$ so that the Ohm law 
\bea
U_2-U_1=R(I_1-I_2)\nn
\eea
will produce linear operators acting on these sets of variables. 

Consider a white vertex in $\Gamma^M$ as the pictures below show.
\begin{figure}[h!]
\label{fig-vert1}
\center
\includegraphics[width=30mm]{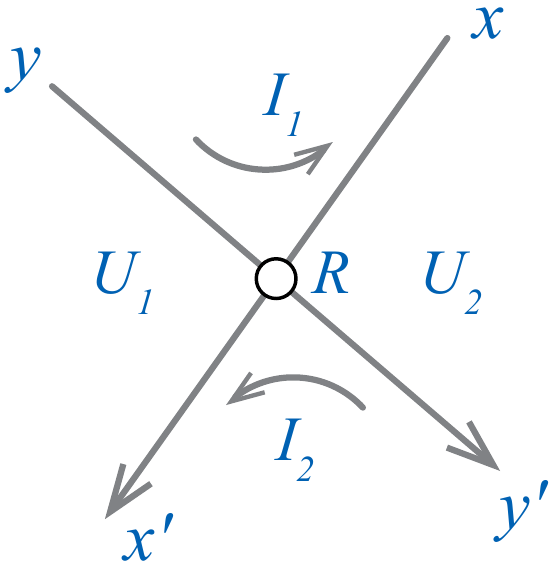}
\caption{White vertex}
\end{figure}

For such a vertex we define the set of edge variables by the equations
\bea
x&=&U_2-I_1;\nn\\
y&=&I_1-U_1;\nn\\
x'&=&I_2-U_1;\nn\\
y'&=&U_2-I_2.\nn
\eea
Rewrite the Ohm law in these variables:
\bea
x+y=R(y-x').\nn
\eea
This equation together with the identity
\bea
x+y=x'+y'\nn
\eea
defines the map $\psi(x,y)=(x',y')$

\bea
\left(
\begin{array}{c}
x'\\
y'
\end{array}\right)=
\left(\begin{array}{cc}
-1/R & 1-1/R\\
1+1/R &1/R
\end{array}\right)
\left(\begin{array}{c}
x\\
y
\end{array}\right).
\eea

which depends of the parameter $R$.
\begin{figure}[h!]
\center
\includegraphics[width=30mm]{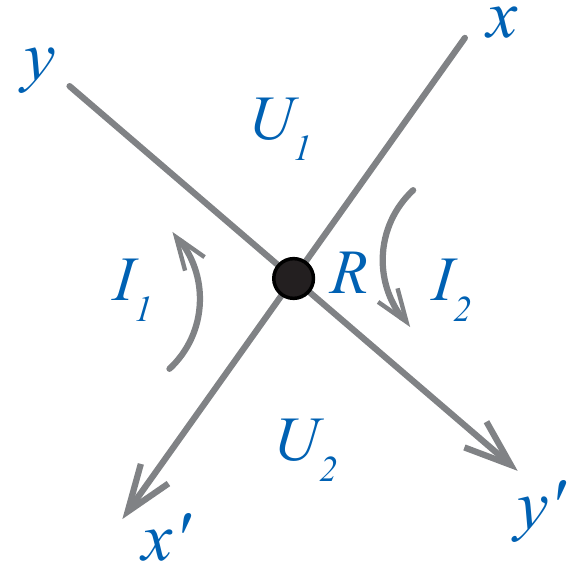}
\caption{Black vertex}
\label{fig-vert2}
\end{figure}
For black vertices we define the set of edge variables by the equations:
\bea
x&=&I_2-U_1;\nn\\
y&=&U_1-I_1;\nn\\
x'&=&U_2-I_1;\nn\\
y'&=&I_2-U_2.\nn
\eea
The Ohm law 
\bea
\frac 1 R(U_2-U_1)=I_2-I_1\nn
\eea
together with the identity
\bea
x+y=x'+y'\nn
\eea 
defines the map  $\phi(x,y)=(x',y')$

\bea
\left(
\begin{array}{c}
x'\\
y'
\end{array}\right)=
\left(\begin{array}{cc}
R & 1+R\\
1-R & -R
\end{array}\right)
\left(\begin{array}{c}
x\\
y
\end{array}\right)
\eea

which depends on the parameter $R$ as well.
These operators are related as follows
\bea
\phi(-\frac1R)=\psi(R)
\eea
We have the following theorem 

\begin{theorem}
The operator $\phi(r)$ satisfies the following  local Yang-Baxter equation.
\bea
\phi_{12}(x_1)\phi_{13}(x_2)\phi_{23}(x_3)=\phi_{23}(x'_3)\phi_{13}(x'_2)\phi_{12}(x'_1)\nn
\eea
where $\phi_{ij}$ is a $3\times 3$ diagonal block matrix those $ij$-block is $\phi$ and the rest is $1$ and
\bea
x'_1&=&\frac{x_1 x_2}{x_1+x_3-x_1 x_2 x_3};\nn\\
x'_2&=&x_1+x_3- x_1 x_2 x_3;\nn\\
x'_3&=&\frac{x_3 x_2}{x_1+x_3-x_1 x_2 x_3}.\nn
\eea
\end{theorem}\label{3.6}
\begin{proof}
The author of \cite{SS}  gives a list of the solutions to the Local Yang-Baxter equation in dimension two. One on the list is the operator $\phi$. 
\end{proof}
The following is the key statement for our approach to the theory of electrical networks.

\begin{corollary} The operator $\phi$ solves the following form of the Local Yang-Baxter equation
\bea
\phi_{12}(R_1)\psi_{13}(R_2)\phi_{23}(R_3)=\psi_{23}(R'_3)\phi_{13}(R'_2)\psi_{12}(R'_1)\nn
\eea
where the parameters $R_i$ and $R'_i$, $i=1,2,3$ are related by the identity:
\label{Star}\[R_j R'_j = R'_1R'_2 + R'_2R'_3 + R'_3R'_1 =\frac{R_1R_2R_3}{R_1 + R_2 + R_3}\]
\end{corollary}
\begin{proof}  Trivial check.
\end{proof}

\subsection{Vertex model of electrical network}
Now we are ready to summarize our construction.
\begin{definition}
For an electrical network $(\Gamma, \gamma)$ define the vertex model associated to it as $(\Gamma^M, {\bf X})$ where the set of matrices ${\bf X}$  is made of the operators $\psi(\gamma(v))$ if the vertex $v$ is white and the operators $\phi(\gamma(v))$
if the vertex $v$ is black.
\end{definition}

\begin{example}
In the picture \ref{fig-network1} the sources are the nodes labelled $1,2,3,5,7$, the rest of the nodes are the sinks. The internal vertices are coloured in white or black,  indicating whether we use the operator $\psi$ or $\phi$ 
to calculate the weights of paths through the internal vertices.
\end{example}
We can introduce as before the set of all vertex models $\frak N(\Gamma^M,{\bf X})$ obtained out of $(\Gamma^M,{\bf X})$ by the Yang-Baxter mutations, except in addition
the Yang-Baxter transformation should turn the black vertices into white ones and vice versa.

Putting together all the pieces we have the following
\begin{theorem} For a given electrical network $(\Gamma,\gamma)$ the  boundary partition function of the associated vertex model is a function on the electrical variety defined by $(\Gamma,\gamma)$.
\end{theorem}
\begin{proof} This is the direct consequence of the fact that $\phi$ and $\psi$ solve the local Yang-Baxter equation \ref{Star}.
\end{proof}

\section{Electrical varieties of standard graphs}
In the section we will study the electrical variety defined by an important set of graphs, the standard graphs, introduced in \cite{CIM}. 
Their  importance is due to the fact that the inverse problem of recovering the conductivities from the response matrix can be solved by the explicit algorithm \cite{CIM}. 
For the electrical variety defined by the standard graph the partition function $M_B$ has
a number of remarkable properties. We will state and prove them in this section.

\subsection{Standard graphs}

The standard graph is defined in \cite{CIM} and is denoted by $\Sigma_n$. The first few of them are figured in \ref{fig:sgraphs} below. 

\begin{figure}[h!]
\center
\includegraphics[width=60mm]{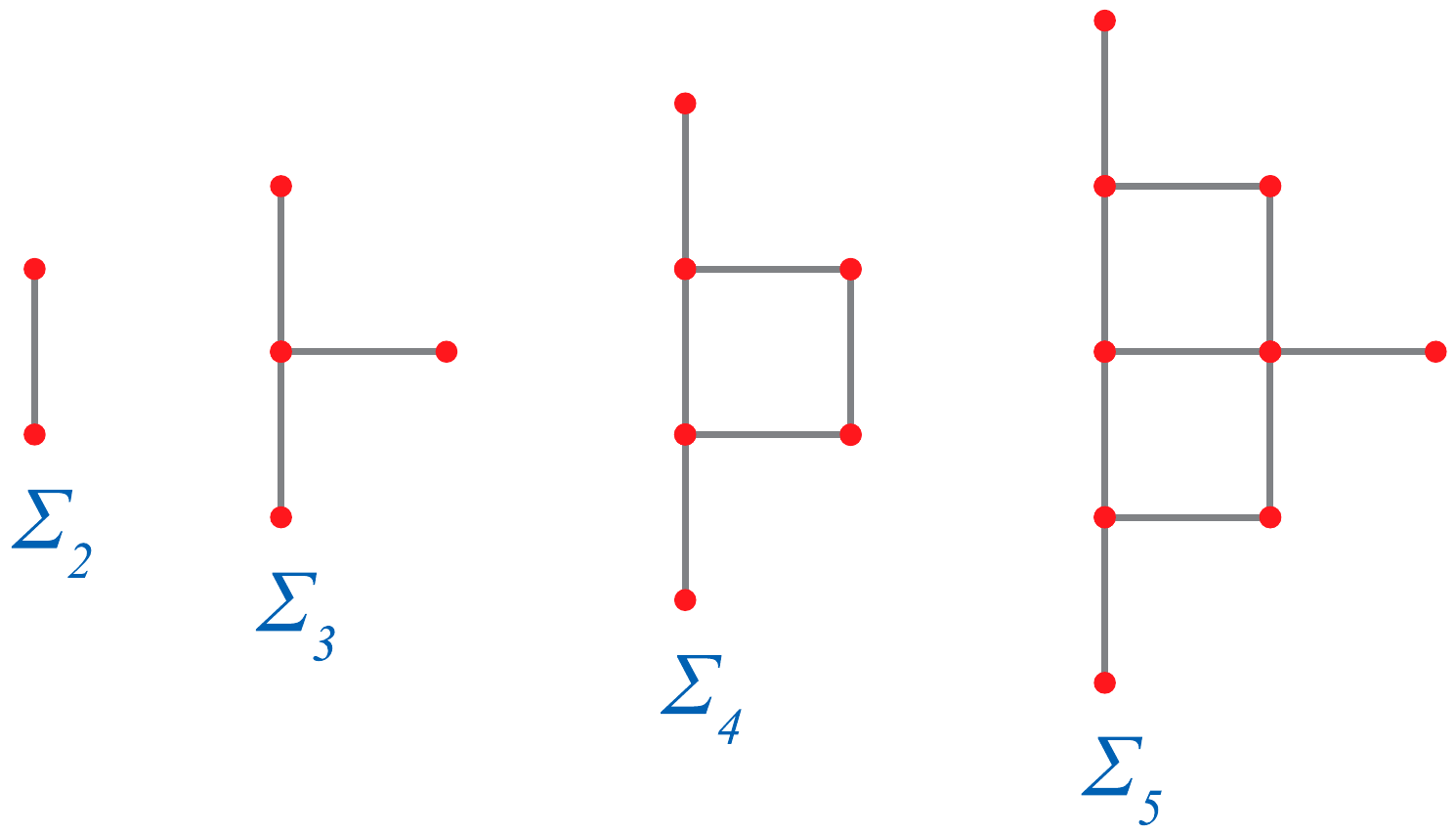}
\caption{Standard graphs}
\label{fig:sgraphs}
\end{figure}
\begin{figure}[h!]
\center
\includegraphics[width=60mm]{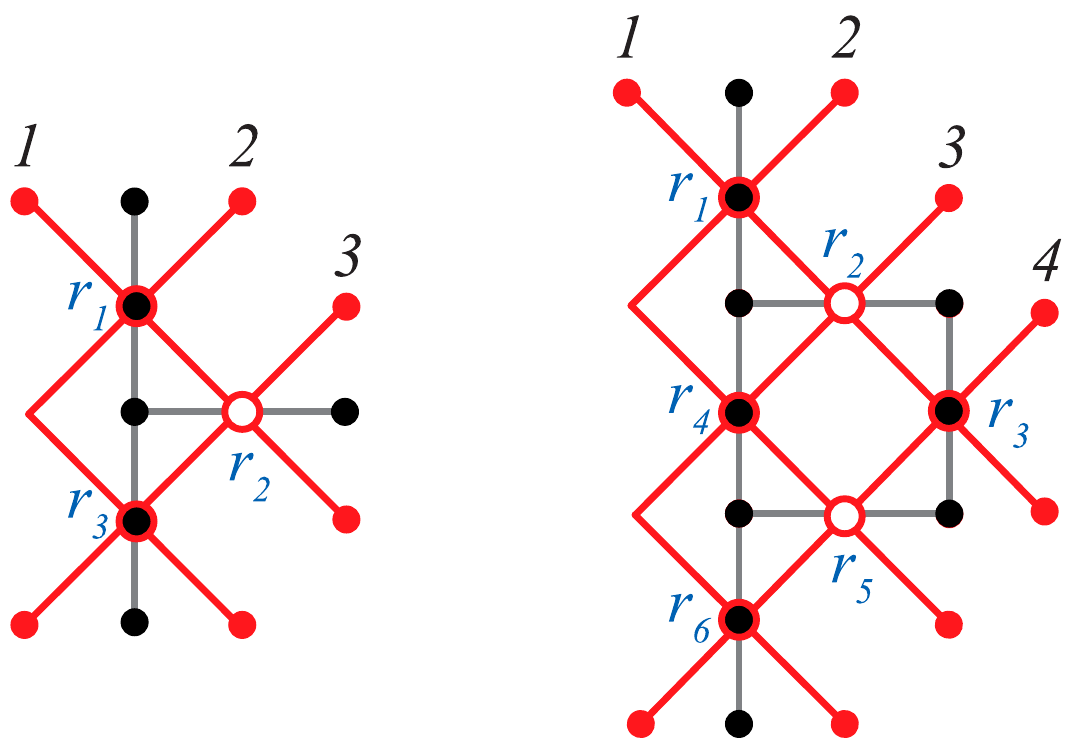}
\caption{Medial graphs}
\label{fig:mgraphs}
\end{figure}

The medial graphs for $n=2$ and $n=3$ will look like the figure \ref{fig:mgraphs} below with the black and white vertices indicated

Convention: we will always label the strands and the vertices of the graph $\Sigma_n^M$ as the picture \ref{fig:mgraphs} shows.  Moreover we label by $n-i+1$ the unlabelled end of the strand labelled by $i$.  These are the sources and the sinks of $\Sigma_n^M$  respectively.
\begin{proposition}
The set  of graphs $\frak N(\Sigma^M_n)$ is exactly the set of pseudo-line arrangements corresponding to reduced words for the longest permutation of $\frak S_n$ modulo 2-moves introduced in \cite{BFZ}.
In particular the graph $\Sigma^M_n$ corresponds to the reduced word $12...(n-1)12...(n-2)12....1$.
\end{proposition} 
\begin{proof} Clear from the figure \ref{fig:mgraphs}.
\end{proof}
\begin{example}
With the labelling of the vertices as in the figure \ref{fig:mgraphs} for $n=2$ we get the pseudo-line arrangement for the word $121$ and for $n=3$ the one for $123121$.
\end{example}

\subsection{The connection between $M_R$ and $M_B$}
\label{MBMR}
Consider the graph $\Sigma_n$.  Here we will prove one of our main results that for any electrical network
on $\Sigma_n$ the response matrix $M_R$ can be recovered from the boundary partition function $M_B$ and visa-versa.

Recall that $M_R$ is a symmetric matrix for which the sum of the entries in each row is zero. On the other hand it is easy to see from the construction of the matrix $M_B$ that the sum of the entries in each row and each column equal to $1$.

First we will describe the connection between the vector of the currents $I=(I_1,...,I_n)$, the vector of the potentials $U=(U_1,...,U_n)$ and the vector of our new face variables $J=(J_1,...,J_n)$ which follows from  
the Ohm law and the Kirchhoff law.

Introduce the matrix $S_n$ of the size $n\times n$
\bea 
S_n=\left(\begin{array}{ccccc}
1 & 0 & \ldots & 0 & -1 \\
-1 & 1 & 0 & \ldots & 0 \\
0 & -1 & 1 & 0 \ldots &0\\
& & \ldots& & \\
0& \ldots & 0 & -1 & 1
\end{array}\right).
\eea
Observe that the matrix $S_n$
has the partial left inverse
\bea 
\tilde S_n=\left(\begin{array}{ccccc}
1 & 0 & \ldots  &  & 0 \\
1 & 1 & 0 & \ldots  & 0\\
1 & 1 & 1 & 0 \ldots & 0 \\
& \ldots &  \ldots&& \\
1& 1&\ldots  & 1 & 0\\
1& 1&\ldots  & 1 & 1
\end{array}\right).
\eea
such that
\bea 
\tilde S_nS_n=\left(\begin{array}{ccccc}
1 & 0 & \ldots  & 0 & -1 \\
0& 1 & 0 \ldots & 0& -1\\
0& 0 & 1&0 \ldots 0 & -1 \\
& \ldots & &  \ldots& \\
0& 0&\ldots & 1 & -1\\
0& 0&\ldots  & 0 & 0
\end{array}\right).
\eea
Introduce also the shuffle matrix $T_{2n}$ of the size $2n\times 2n$ which moves the components $U$ of the vector $(J,U)$ through the components $J$ putting $U_i$ in front of $J_i$ for all $1\leq i\leq n$.
\begin{example}
Here are the matrices $S_3$ and $T_6$
\bea 
S_3=\left(\begin{array}{ccc}
1 &  &   -1 \\
-1 & 1 &  \\
 & -1 & 1  
\end{array}\right),
\eea
\bea
T_6=
\left(\begin{array}{cccccc}
& & & 1 & &\\
1 & & & & & \\
& & & & 1 & \\
& 1 & & & & \\
& & & & & 1 \\
& & 1 & & &
\end{array}\right).
\eea
\end{example}
The labelling of the input and the output vertices in the medial graph of $\Sigma_n$ is given  as follows. The first $n$ components of the vector
\bea
S_{2n}T_{2n}\left(\begin{array}{c}
J \\
U 
\end{array}\right)
\eea
label the inputs and the other $n$ components label the outputs.

\begin{figure}[h!]
\center
\includegraphics[width=70mm]{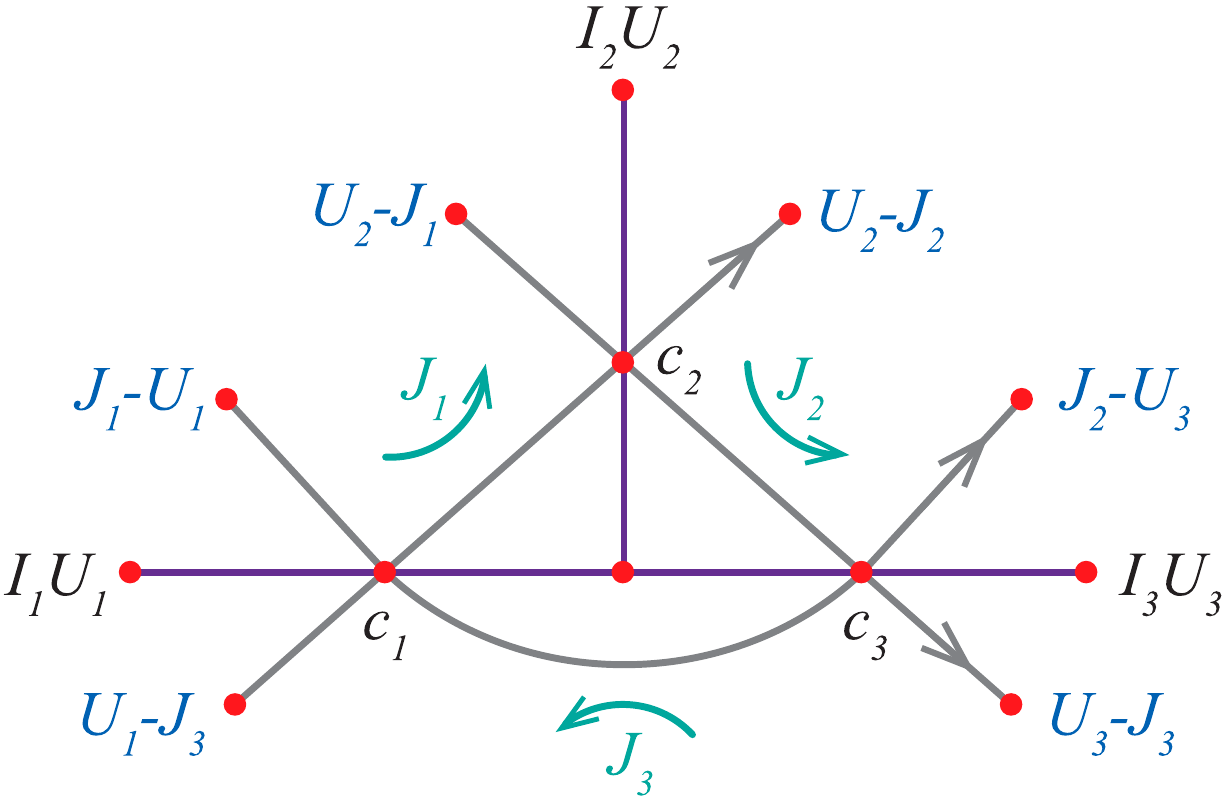}
\caption{Edge variables}
\label{fig:edge-var}
\end{figure}

The picture  \ref{fig:edge-var}  illustrates it for $n=2$.
The variables $I$, $J$, and $U$ are connected by the following equations.
First is the definition of the response matrix of our electrical network
\[I=M_RU.\]
The other is our definition of the face variables $J$
\bea
I=S_n J.\nn
\eea
Finally the definition of our partition function gives
\bea
(M_B,{\bf Id})S_{2n}T_{2n}\label{repr1}
\left(\begin{array}{c}
J \\
U 
\end{array}\right)=0.
\eea
where ${\bf Id}$ is the identity matrix of the size $n\times n$.
\begin{example}
For $n=2$ it looks like
\bea
\label{2-nd-repr}
M_B
\left(\begin{array}{c}
U_1-J_3 \\
J_1-U_1\\
U_2-J_1\\
\end{array}\right)=
\left(\begin{array}{c}
U_2-J_2\\
J_2-U_3 \\
U_3-J_3

\end{array}\right),
\eea
\end{example}
The above leads to the key observation
\begin{lemma}\label{grassmanian}
The matrices
\bea
W_1=(S_n, M_R)\nn
\eea
and
\bea
W_2=(M_B,{\bf Id}_n)S_{2n} T_{2n}\nn
\eea
define the same point in the Grassmanian $Gr(n-1,2n)$.
\end{lemma}
\begin{proof} The proof is obvious since both matrices when considered as linear operators have the same kernel. Since the matrix $S_n$ has the partial left inverse $\tilde S_n$ and the matrix $M_R$ is singular, the rank of the matrix 
$W_1$ is equal to $n-1$ and  hence this matrix defines a point in $G(n-1,2n)$ as claimed. This says of course that the rank of the matrix $W_2$ must be $n-1$. Indeed although 
the matrix $M_B$  is non-degenerate as we pointed out above, the sum of the entries in each column of $M_B$ is equal to $1$, hence 
the matrix $(M_B,{\bf Id}_n)S_{2n}$ has the rank less than $n$. But then it must be equal to $n-1$ to match the rank of $W_1$.
\end{proof}
Let ${\bf Id}_0$ be the $(n-1)\times n$ matrix obtained out of the identity matrix ${\bf Id}_n$ by crossing out the last row and ${\bf Id}_1$ be the $2n\times 2n-1$ matrix obtained out of the identity matrix ${\bf Id}_{2n}$
by crossing out the last column. Denote by $M_0$ the left $n-1\times n-1$ submatrix of the matrix ${\bf Id}_0W_2$
and denote by $M_1$ the right $n-1\times n-1$ submatrix of the matrix ${\bf Id}_0W_1T_{2n}\tilde S_{2n}{\bf Id}_1$.
\begin{theorem}
The matrix ${\bf Id}_0\tilde S_n M_R$ is the right $(n-1)\times n$ (block)  submatrix of the matrix
$$
M^{-1}_0{\bf Id}_0W_2
$$
The matrix ${\bf Id}_0M_B$ is the left $(n-1)\times n$ submatrix of the matrix 
$$M^{-1}_1{\bf Id}_0W_1T_{2n}\tilde S_{2n}$$.
\end{theorem}\label{response}
\begin{proof}
Multiplying $W_1$ by $\tilde S_n$ on the left and crossing out the last row we get the matrix on the form $({\bf Id}_{n-1}, {\bf -1},\tilde M_R)$, where ${\bf -1}$ is  the $(n-1)$ dimensional column vector whose components are all equal to $-1$.
Therefore by \ref{grassmanian} the $M_0$ is invertible and the first statement of the theorem follows.

The group $GL(2n)$ action on the vector space $\Bbb C^{2n}$ induces the action of $GL(2n)$ on $G(n-1,2n)$. Therefore the matrices $W_1T_{2n}\tilde S_{2n}$ and $W_2S_{2n}\tilde S_{2n}$ still
represent the same point in $G(n-1,2n)$. The matrix ${\bf Id}_0W_2S_{2n}\tilde S_{2n}$ has the form $({\bf Id}_0M_B,{\bf Id}_{n-1},-{\bf 2})$, where ${\bf -2}$ is the $(n-1)$ dimensional column vector with all the components equal to $-2$.
Again by \ref{grassmanian} $M_1$ is invertible and the statement follows.
\end{proof}
We finish this subsection with a couple of remarks.
\begin{remark}
Our theorem  should be helpful for calculating the response matrix. Being the Schur complement the response  matrix requires inverting $(n(n-1)/2)\times(n(n-1)/2)$ matrix. 
The above theorem says one needs to invert the $(n-1)\times(n-1)$ matrix $M_{B,0}^J$ to calculate the response matrix.
\end{remark}
\begin{remark} According to \cite{CIM} the response matrix of an electrical network has some interesting positivity properties. Namely all the so called {\it circular minors} have the same sign. The results we have presented in this section suggest that the boundary partition function associated to an electric network also has some special positivity properties. We plan to study this in a future publication.
\end{remark}
The appearance of the Grassmanian $G(n-1,2n)$ we saw above should be compared to the way it appears in \cite{Lam}. This could lead to identifying precisely the place of our complete electrical invariant $M_B$ in the earlier work on the electrical varieties.

\subsection{Symplectic group and Temperley-Lieb algebra}
Recall that the boundary partition function $M_B$ is the product of the operators $\phi(r)$, the solutions for the local Yang-Baxter equation. The matrix $M_B$ for the standard graph  $\Sigma_2$ \ref{MBMR} 
for example is the following product
\bea
M_B=\phi_{23}(-1/r_3)\phi_{13}(r_2)\phi_{12}(-1/r_1).
\eea 
For the graph $\Sigma_k$, $k>2$ the matrix $M_B$ takes the form
\bea
M_B=\prod_{ i<j} \phi_{ij}((-1)^{i+j}r_{ij}^{(-1)^{i+j}})
\eea
where the product is over the pairs $(ij)$ such that $i<j$ and the order of the factors is defined by the lexicographic order on the pairs $(ij)$.

To make connection with the Lusztig varieties it is convenient to consider the alternative version for the boundary partition function.
There are two forms of the Yang-Baxter equation
\bea
R_{12} R_{13} R_{23}=R_{23} R_{13} R_{12},\nn
\eea
and
\bea
\check{R}_{12} \check{R}_{23} \check{R}_{12}=\check{R}_{23} \check{R}_{12} \check{R}_{23},\nn
\eea
where 
\bea
\tilde{R}=P R,\nn
\eea\label{check}
and $P$ is the permutation operator.

Introduce the operators $\check{\phi}_{ij}=P_{ij} \phi_{ij}$ and define as before the
product
\bea
M_B=\prod_{ i<j} P_{ij}\check\phi_{ij}((-1)^{i+j}r_{ij}^{(-1)^{i+j}}).\nn
\eea
over the pairs $(ij)$ in the lexicographic order.
Moving all the permutation operators to the left we obtain
\bea
M_B=\prod_{ i<j} P_{ij}\prod_{ i<j} \check\phi_{j-i\,j-i+1}((-1)^{i+j}r_{ij}^{(-1)^{i+j}}).\nn
\eea
Define the matrix $\check M_{B}$ by the following formula
\bea\label{sympl}
\check M_{B}=\prod_{ i<j} \check\phi_{j-i\,j-i+1}((-1)^{i+j}r_{ij}^{(-1)^{i+j}}).
\eea
\begin{example}
For the graph $\Sigma_2$ this matrix takes the form:
\bea
\check M_{B}=\check\phi_{12}(-1/r_3)\check\phi_{23}(r_2)\check\phi_{12}(-1/r_1).
\eea
\end{example}

We put the above calculations into the following
\begin{proposition}\label{cphi}
$M_B=\omega_0\check M_B$, where $\omega_0$ is the longest element of the symmetric group represented by the matrix
\bea
\left(\begin{array}{ccccc}
0 & \ldots  &  & 0 & 1 \\
0 & \ldots &  & 1 & 0\\
&  &  &  \ldots &  \\
0 & 1 & \ldots & &  0\\
1& 0 &\ldots  & 0 & 0
\end{array}\right).
\eea
\end{proposition}

\begin{theorem} The boundary partition function $\check M_B$ of an electrical variety preserves the bilinear form
\bea
\Omega_n=\sum_{1\le i<j\le n} x_i\wedge x_{j}
\eea
which is symplectic for even $n$.
\end{theorem}
\begin{proof}   The operator $\check\phi_{i\,i+1}(r)$ belongs to the symplectic group $Sp(n)$ for $1\leq i\leq n-1$ and even $n$:
\bea
\check\phi_{i\,i+1}(r)^T \Omega_n \check\phi_{i\,i+1}(r)=\Omega_n.
\eea
The form $\Omega_n$ and the operator $\phi_{i\,i+1}(r)$ are given by the following matrices:
\bea
\Omega_n&=&\sum_{i<j} e_{ij}-\sum_{i>j}e_{ij};\nn\eea
\bea\check\phi_{i\,i+1}(r)&=&Id+r(-e_i+e_{i+1})(e_i+e_{i+1})^T=Id+r a_i.\nn
\eea
where $e_{ij}$ is the matrix unit and $e_i$ is the standard basis vector.
It is clear that:
\bea
\Omega_n a_j&=&-(e_j+e_{j+1})(e_j+e_{j+1})^T;\nn\\
a_j^T \Omega_n&=&(e_j+e_{j+1})(e_j+e_{j+1})^T;\nn\\
a_j^T \Omega_n a_j&=&0.\nn
\eea
Therefore
\bea
\check M_B^T \Omega_n \check M_B=\Omega_n.
\eea

After change of the basis 
\bea
v_i&=&v'_i-v'_{i+1}, \,\,0<i<n\nn\\
v_n&=&v'_n\nn
\eea 
the form $\Omega_n$  will be given by the matrix
\bea
\left(\begin{array}{ccccc}
0 & 1  & 0  & \ldots & 0 \\
-1 & 0 & 1 & \ldots & 0\\
 0 & -1 & 0 &  & \ldots \\
\ldots &  & \ldots & 0 &  1\\
0& 0 &\ldots  & -1& 0
\end{array}\right).
\eea
which is known to be symplectic for even $n$.
\end{proof}
\begin{remark} The proof above works for any electrical network with the even number of nodes $n$, in other words the partition function of any such a vertex model belongs to the symplectic group $Sp(n)$.
\end{remark}

When $n$ is odd the partition function $M_B$ also has interesting properties, namely its determinant is equal to $1$, hence it belongs to the group $Sl(n)$, and the sum of the matrix elements in every columns is equal to $1$ as we already have mentioned.

Recall that the operator $\check\phi(r)$ satisfies the following form of the local Yang-Baxter equation.
\[( \check\phi_{12}(r_3)) (\check\phi_{23}(r_2))(\check\phi_{12}(r_1))=(\check\phi_{23}(r'_3))(\check\phi_{12}(r'_2)) (\check\phi_{23}(r'_1))\]
\bea\label{LYB}
r'_1&=&\frac{r_3r_2}{r_1+r_3-r_1r_2r_3};\nn\\
r'_2&=&r_1+r_3- r_1r_2r_3;\nn\\
r'_3&=&\frac{r_1r_2}{r_1+r_3-r_1r_2r_3}.\nn
\eea

Following \cite{BFZ} introduce an associative algebra  with the generators $u_1,...,u_{n-1}$ and with the relations defined by the following equations satisfied for arbitrary $r_1,r_2,r_3$:
\[(1+r_1 u_i)(1+r_2 u_j)=(1+r_2 u_j)(1+r_1 u_i), \]
if $|i-j|\geq 2$,
\[(1+r_1 u_i)(1+r_2 u_j)(1+r_3 u_i)=\]\[(1+\frac{ r_2 r_3}{r_1+r_3-r_1r_2r_3}u_j)(1+(r_1+r_3-r_1r_2r_3) u_i)(1+\frac{ r_1 r_2}{r_1+r_3-r_1r_2r_3}u_j)\]\\
if $|i-j|=1$. 

Comparing the coefficients of
identical monomials  in $r_i$'s on both sides we obtain
\[u_iu_j=u_ju_i,\,\,\,|i-j|>1\]
\[u_i^2=0\]
\[u_iu_ju_i=-u_i,\,\,|i-j|=1\]
This makes our algebra the Temperley-Lieb algebra specialized at the root of unity $q=\bf{ i}$. We will denote it by $TL(0)$.

Define the Temperley-Lieb algebra with coefficients in the polynomial ring ${\bf r}:=\Bbb C[r_1,...,r_m]$ as $TL(0)\otimes_{\Bbb C} {\bf r}$. We will be denoting it by the same letter $TL(0)$.
\begin{theorem}
 The following formula defines the map from the electrical variety defined by the standard graph $\frak T(\Sigma_n)$ to $TL(0)$:
\[\phi:(r_1,...r_m)\rightarrow(1+r'_1u_1)(1+r'_2u_2)...(1+r'_{n-1}u_{n-1})(1+r'_nu_1)....(1+r'_mu_1)\]
where 
 \[
    r'_i=\left\{
                \begin{array}{ll}
                  r_i, \,\,i\text  { is black} \\
                  -r_i^{-1},\,\,i\,\,\text  {is white}
                \end{array}
              \right.
  \]
  \end{theorem}
\begin{remark}
The Lusztig solution to the Local Yang-Baxter equation used in \cite{BFZ} is related in the same way to the nil-Temperley-Lieb algebra, a degeneration of our Temperley-Lieb algebra.
\end{remark}
\begin{remark}
It turns out that the representation of the Temperley-Lieb algebra $TL(0)$ at a root of unity $q={\bf i}$ related to the Ohm-Yang-Baxter operator $\phi$ appears in the studying of the $XXZ$ model in statistical physics \cite{Azat}. 
The explicit formula for the generators $u_i$ of $TL(0)$ in terms of the symplectic algebra generators is given in \cite{Azat} $6.11, 6.12$.
\end{remark}
\begin{remark} The results in this section should be compared to  the electrical Lie group defined in  \cite{LP}.
\end{remark}
\subsection{The solution of the inverse problem}
 As we mentioned earlier the inverse problem for an electrical network $\Gamma,\gamma$ is recovering the values of the conductivity function from the known response matrix $M_R$. It can not be solved in general, however in the case when $\Gamma$ is the standard graph there is an explicit algorithm which solves it \cite{CIM}. The algorithm is based on the boundary properties of discrete harmonic functions. 

In our picture the boundary partition function plays the role of the response matrix and the inverse problem for it is a an interesting task. Finding the solution of the inverse problem for the Lusztig varieties was an important 
achievement in \cite{BFZ}. 

We present here a simple algorithm for finding the values of the function $\gamma$ out of the matrix $M_B$ for an electrical network defined on standard graph $\Sigma_n$:

Let the interior vertex $v_{ij}$ be the intersection of the strands started at the sources labelled $i$ and $j$, $i<j$. By the way we label the boundary vertices, these strands arrive to the sinks labelled $n-i+1$ and $n-j+1$ respectively see figure (\ref{fig:mgraphs}). There is only one path connecting the boundary vertices labeled $j$ and $n-i+1$ which contains the vertex $v_{ij}$, it is the path which runs along these two strands only. The rest of the paths 
which connect these two vertices must be strictly on the right of it as the figure (\ref{fig:mgraphs}) shows. Therefore in the sum which calculates the matrix entry $(M_B)_{j,n-i+1}$ \ref{partition function} there is only one summand which contains the factor $\gamma(v_{ij})$, the rest will be monomials in $\gamma(v_{kl})$ with either $j<k$ or $n-i+1<l$ or both. This provides the step of the induction. Since $(M_B)_{nn}=1-\gamma(v_{nn})$ we obtain the algorithm.

The result of the previous sections together with the above algorithm allow to prove 
\begin{corollary} The boundary partition function defines an embedding of the variety $\frak T$ defined by the standard graph $\Sigma_n$ into the symplectic group $Sp(2n)$.
\end{corollary}

\begin{remark} Combining this algorithm for the matrix $M_B$ with the theorem \ref{response} we obtain a new solution for the inverse problem for electrical networks on the standard graphs. 
Indeed given the response matrix of such a network we can calculate out of it the boundary partition function $M_B$ using \ref{response} and then calculate the conductivities using our algorithm.
\end{remark}

\section{Electrical varieties as deformation of Lusztig varieties}
In this section we will present several properties of the electrical varieties which show that the theory of the electrical varieties is an interesting deformation of the theory of the Lusztig varieties.
For this we introduce a slight modification of the electrical variety defined by the standard graph. We simply ignore the colours on the vertices of the graph $\Gamma^M$ and use the collection ${\bf X}$ of matrices made out of the operators $\check\phi(r)$ defined earlier \ref{check}. 
We will denote the resulting variety $\frak L_1$. It is in fact isomorphic to $\frak T$, but it is more convenient for studying the connection to the variety $\frak L$.

\subsection{The boundary partition functions for the Lusztig and the electrical varieties}
 To put the electrical varieties and the Lusztig varieties on the same footing we will present the latter as a vertex integrable statistical model. In this language the Lusztig variety is a pair $(\Sigma_n^M, {\bf X})$, 
with the collection of matrices ${\bf X}$ made out of the operators
\bea\label{lus}
\varphi^L(t)=\left(
\begin{array}{cc}
1 & t\\
0 & 1
\end{array}
\right).
\eea

The operator $\varphi_L(t)$ satisfies the following  local Yang-Baxter equation.
\[ \varphi_{12}^L(t_3)\,\varphi_{23}^L (t_2)\,\varphi_{12}^L(t_1)=\varphi_{23}^L(t'_3)\,\varphi_{12}^L(t'_2)\,\varphi_{23}^L(t'_1)\]
where
\bea
t'_1&=&\frac{t_2 t_3}{t_1+t_3};\nn\\
t'_2&=&t_1+t_3;\nn\\
t'_3&=&\frac{t_1t_2}{t_1+t_3}.\nn
\eea
\begin{remark} 
Since we use the operators $\check\phi(r)$ and $\varphi^L(t)$ we must {\bf reverse} the order of the sinks of the graph $\Sigma_n^M$ as opposed to our earlier convention, see \ref{cphi}.
\end{remark}
As such this variety is a vertex model in our definition and it has the boundary partition function. It is represented by an upper triangular unipotent matrix as it should be according to \cite{BFZ}. 

\begin{theorem} Denote by $M_{\frak L}$ and $M_{\frak L_1}$ the boundary partition functions for the Lusztig and the electrical varieties defined by the standard graph. The matrix elements of $M_{\frak L_1}$ are polynomials in the variables $\gamma(v)$. Denote by $M^0_{\frak L_1}$ the upper triangular matrix whose matrix coefficients are the homogeneous parts of the smallest degree of the coefficients of the upper triangular submatrix of  $M_{\frak L_1}$. Then $M_{\frak L}=M^0_{\frak L_1}$.
\end{theorem}\label{LEl}

\begin{proof}
Call a path in the standard graph {\it non-decreasing} if it does not contain a slant line interval going from the level $i$ to the level $i-1$ in the pseudo-line presentation. 
The smallest  degree homogeneous component of the weight of a path if it connects the vertices $v_i$ and $v_j$, $i\leq j$, is $|i-j|$.
It is clear that only  the non-decreasing paths have the smallest degree component in the weight polynomial.

Comparing the weights of a non-decreasing path defined by the matrix $\phi$ and the weight of this path calculated using the weights defined by the matrix elements of the matrix (\ref{lus})
we see that for a standard graph the smallest degree component in the weight polynomial of the former is the weight of the latter. 
But the non-decreasing paths calculate the upper triangular part of the boundary partition function for the electrical variety on the one hand and the boundary partition function of the Lusztig variety on the other hand.
\end{proof}

\subsection{The cluster coordinates}

The important observation made in \cite{BFZ} about the algebra of functions on the Lusztig variety is the following: introduce the set of variables $M_L$, where $L\subset [1,n]$.
Let $h=(h_1, ..., h_m)$, where $1\leq h_k \leq n-1$, be a reduced word for the longest permutation $w_0$ labelling a chart ${\bf t}^h=(t^h_1,\ldots,t^h_m)$ in the variety $\frak L$. For each entry $h_k$ of $h$ define a Chamber set $L\subset [1, ..., n]$ for $h$, and two integers $i$ and $j$ by
\[L=s_{h_m }\ldots s_{h_{k+1}}(\{1, ..., h_{k-1}\})\]
\[ i=s_{h_m}\ldots s_{h_{k+1}} (h_k) \]
 \[ j=s_{h_m} \ldots s_{h_{k+1}}(h_{k+1})\]
For example, for $h=213231$ and $k=3$, we obtain $(i, j)=(1, 3)$ and $ L=[2, 4]$. 
\begin{theorem} \cite{BFZ} Let  $t^h_k$ be defined by the Chamber Ansatz substitution
\[t^h_ k=\frac{M_LM_{L\cup \{i, j\}}}{ M_{L\cup \{i\}}M_{L\cup \{j\}}}\]
where $i$, $j$, $L$ are defined above.
Then the point ${\bf t}^h$ belongs to the Lusztig variety $\frak L$ if and only if the variables $M_L$ satisfy the following relation
\[M_{L\cup \{i, k\}}M_{L\cup \{j\}}=M_{L\cup \{i, j\}}M_{L\cup \{k\}}+M_{L\cup \{j,k\}}M_{L\cup \{i\}} \]
 whenever $i<j<k$ and $L\cap \{i, j,k\}=\emptyset$. 
\end{theorem}
In other words, the Chamber Ansatz translates the gluing functions for the charts of $\frak L$  into the 3-term classical Pl\"ucker relations.
In fact there is the inverse map.
Let $M(L)$ be the subset of $\{M_L\}$ formed by those tuples $M=(M_L)$ so that, in addition to the above relations, it satisfies the normalization condition
$M_{\emptyset}=1,\,\, M_{[1, b]}=1, b=1, ..., n. $
\begin{theorem}\cite{BFZ} The restriction of the Chamber Ansatz map $M$ onto $M(L)$ is a bijection between $M(L)$ and the Lusztig variety $\frak L$. The inverse bijection 
between ${\frak L}$ and $M(J)$ is  given
\[M_J=\prod_{k:j\in J,\, i<j, \,i\not\in J} (t^h_k)^{-1}\]
whenever $J$ is a chamber set for $\bf h$. 
\end{theorem}

\begin{example} Let us illustrate these theorems in the case of the graph $\Sigma_2^M$. The only other graph we can obtain out of it by the Yang-Baxter mutation corresponds to the word $h=(212)$. The chamber sets are indicated in the picture \ref{fig:chamber}.
\begin{figure}[h!]
\center
\includegraphics[width=100mm]{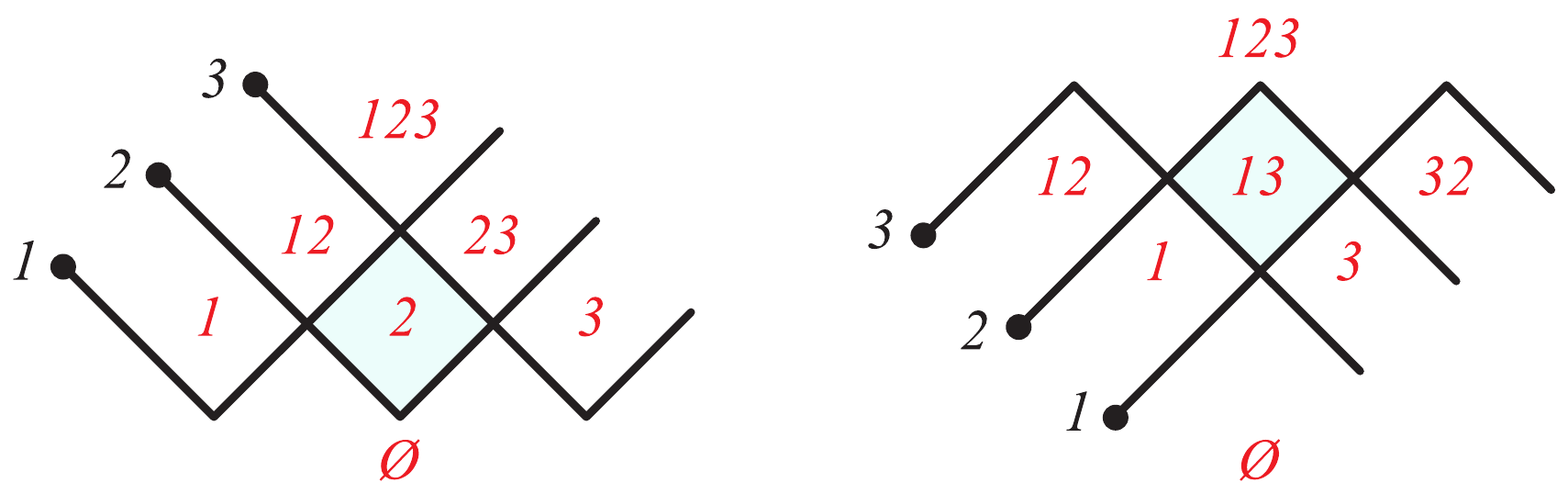}
\caption{Chamber Ansatz}
\label{fig:chamber}
\end{figure}
The formulas for the Chamber Ansatz and the inverse map are given below
\[t_1=\frac{M_{\emptyset}M_{12}}{M_1M_2}\,\,\,\,M_2=\frac1{t_1}\,\,\,\,t_1'=\frac{M_{1}M_{123}}{M_{12}M_{13}}\,\,\,\,M_{13}=\frac1{t'_1}\]
\[t_2=\frac{M_{2}M_{123}}{M_{12}M_{23}}\,\,\,\,M_{23}=\frac1{t_1t_2}\,\,\,\,t_2'=\frac{M_{\emptyset}M_{13}}{M_{1}M_{3}}\,\,\,\,M_3=\frac1{t'_1t'_2}\]
\[t_3=\frac{M_{\emptyset}M_{23}}{M_2M_3}\,\,\,\,M_3=\frac1{t_2t_3}\,\,\,\,t_3'=\frac{M_{3}M_{123}}{M_{13}M_{23}}\,\,\,\,M_{23}=\frac1{t'_2t'_3}\]
The point is that $M_3$ and $M_{23}$ are chamber sets in both of the charts but because $t_2t_3=t'_1t'_2$ and $t_1t_2=t'_2t'_3$ they are defined unambiguously.

The 3-term relation takes the form:
\[M_2M_{13}=M_3M_{12}+M_1M_{23}\]
\end{example}
Quite remarkably similar statements hold  for the electrical variety. It was first observed in \cite{KW}.
Mimicking the Chamber ansatz from \cite{BFZ} introduce the set of variables $\widehat M_L$, where $L\subset [1,n]$. For any $h\in \frak N (\Sigma^M_k)$, the appropriate chart ${\bf r}^h=(r^h_1,\ldots, r^h_m)$ and $l\in [1,m]$ set
\[r^h_l=\frac{\widehat M_L\widehat  M_{L\cup \{i, j\}} }{\widehat M_{L\cup i}\widehat M_{L\cup j}}\]
These variables are call the {\bf B}-variables in \cite{KW}.
This way to assign the coordinates to the vertices edges and the faces of the local partition of the surface defined by the graph on it was also used in \cite{GK} to state the cluster algebra nature of the star-triangular transformation.

The next theorem shows that the three term cluster relation between $M_L$ in the algebra of functions of $\frak L$ deforms to a four term cluster relation between the variables $\widehat M_L$ but the Chamber Ansatz is still invertible.
 \begin{theorem}\cite{KW}
The point $(r^h_i)$ belongs to $\frak L_1$  if and only if the variables $\widehat M_L$ satisfy 
\[\widehat M_{L\cup \{i, k\}}\widehat M_{L\cup \{j\}}=\widehat M_{L\cup \{i, j\}}\widehat M_{L\cup \{k\}}+\widehat M_{L\cup \{j,k\}}\widehat M_{L\cup \{i\}}+\widehat M_L\widehat M_{L\cup\{i,j,k\}} \]
 whenever $i<j<k$ and $L\cap \{i, j,k\}=\emptyset$. 
 
 Moreover the restriction of the Chamber Ansatz map to $\widehat M(L)$defined in the same way as $M(L)$ is a bijection between $\widehat M(L)$ and the variety $\frak L_1$. The inverse bijection 
${\frak L_1}$ to $\widehat M(L)$ is  given
\[\widehat M_J=\prod_{k:j\in J,\, i<j, \,i\not\in J}(r^h_k)^{-1}\]
whenever $J$ is a chamber set for $\bf h$ and the product is over all $k$ such that $i\not\in J$, $j\in J$. 
 \end{theorem}
\begin{proof}
The first statement is proved by a direct calculation.

As for the second statement, the proof from \cite{BFZ} uses only the following property of the transition functions: 
if two charts are related by the elementary transformation
$t_i,t_j,t_k\rightarrow t'_i,t'_j,t'_k$ then $t_it_j=t'_kt'_j$ and  $t_kt_j=t'_it'_j$. This holds for the transition maps of the electrical variety \ref{LYB} as well.
\end{proof}
\begin{example}
For the electrical variety associated to the same graph $\Sigma_2^M$ the Chamber Ansatz formulas are the same as for the Lusztig variety but the 3-term relation deforms to the 4-term relation:
\[M_2M_{13}=M_3M_{12}+M_1M_{23}+M_{\emptyset}M_{123}\]
\end{example}
This theorem suggests that the theory of the electrical varieties may provide a non-trivial deformation of the cluster algebra structure introduced in \cite{FZ}. 

In the case of the Lusztig variety the variables $M_L$ have a nice interpretation as the minors of a matrix closely related to the partition function of the Lusztig variety \cite{BFZ}. The interpretation of these variables in the case of the electrical variety is not known at the moment as far as we understand, however the theorem \ref{LEl} maybe be helpful in finding it. We plan to study it in a future publication.

\section{Discrete integrable dynamics of electrical varieties}

It is natural to study the transformations of the electrical varieties for which the behaviour of the response matrix is under control. The list of such is known (see for example \cite{CdV}). The most interesting for us are the star-triangle transformation and the transformation adjoining an edge to the graph. In terms of the medial graph this is equivalent to the (local) Yang-Baxter transformation and the transformation of adding a crossing to the medial graph.
\subsection{General scheme}
There is a class of discrete systems of the Toda type related to the Lusztig variety. These were considered for example in \cite{Yam} and are related to the box-ball system, the Painlev\'e IV equation and many other dynamical systems. 
The machinery of the electrical varieties offers a natural generalization of these. We consider the discrete system produced by commuting 
actions of a pair of the symmetric groups each defined by the above transformations of the electrical variety. This system was elaborated in a series of papers summarized by \cite{LP2}.  We include our calculations here to demonstrate the similarity with the result of \cite{Yam}. These actions are the deformations of the actions studied in \cite{Yam} and in the appropriate limit recover the action considered in \cite{Yam}, see also \cite{LP1}, \cite{ball-box}.

\begin{figure}[h!]
\center
\includegraphics[width=120mm]{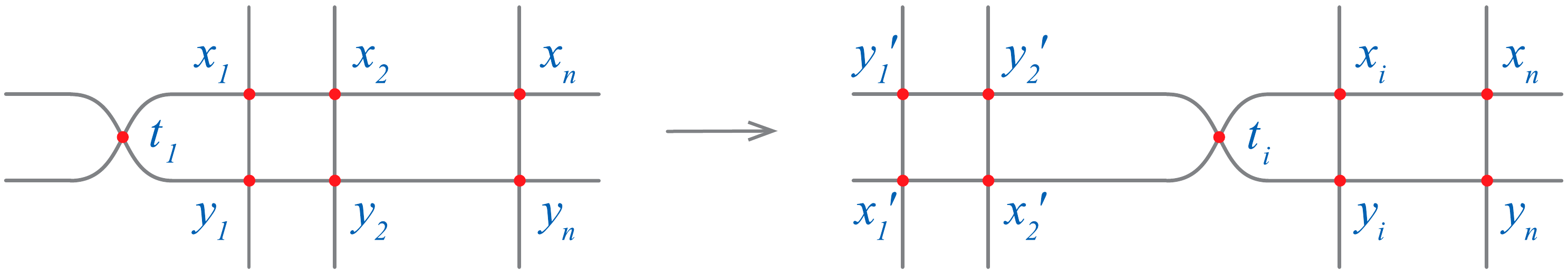}
\caption{Generators $\pi$'s}
\label{symm1}
\end{figure}

Consider the graph whose medial graph is a ladder made out of two horizontal lines and a bunch of vertical lines as in the picture below. Following \cite{LP1} add a crossing to this graph with the operator $\phi(t)$ attached to the new vertex as shown on the figure \ref{symm1}.

For a particular choice of $t$ the rectangular network transforms to the rectangular network with the new resistances attached to the vertices sitting in the intersection points of the horizontal and the vertical lines. 

\begin{figure}[h!]
\center
\includegraphics[width=80mm]{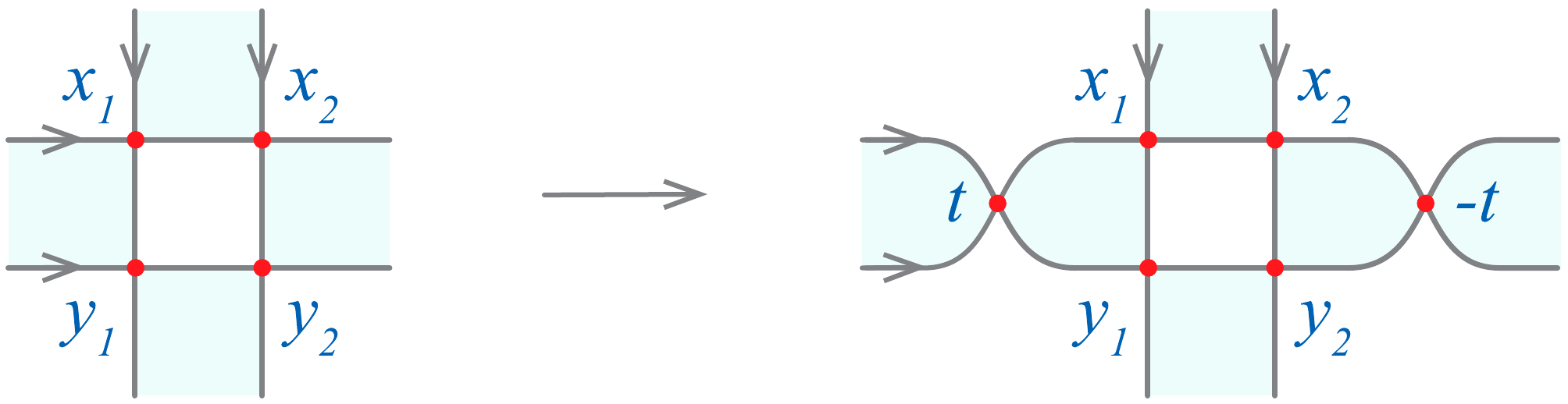}
\caption{2x2 network}
\label{symm2}
\end{figure}

\begin{figure}[h!]
\center
\includegraphics[width=25mm]{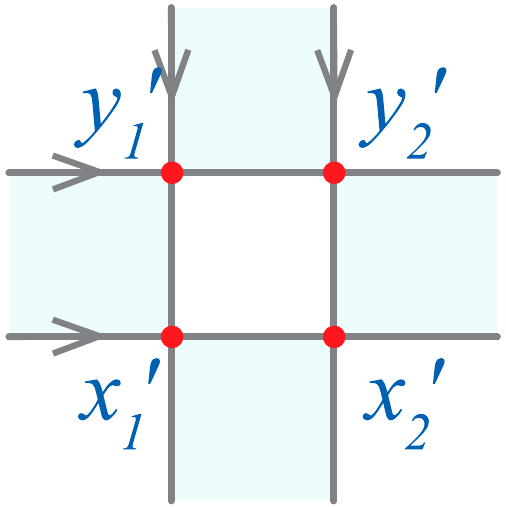}
\caption{Equivalent network}
\label{symm3}
\end{figure}

For the network as on the figure \ref{symm2} and the following choice of $t$
\bea
t=\frac{x_1 x_2 - y_1 y_2} {x_1 + y_2 + x_1 x_2 y_2 + x_1 y_1 y_2}
\eea 
the network on the figure  \ref{symm2} is equivalent to the network on the figure \ref{symm3}.
The transformation of the resistances is given by:
\bea
x_1'=x_1 \mu ; \qquad y_1'=y_1 \mu^{-1}; \qquad x_2'=x_2 \mu^{-1}; \qquad y_2'=y_2 \mu;\nn
\eea
and
\bea
\mu=(x_1+y_2+x_1 y_2 (x_2+y_1))/(x_2+y_1+x_2 y_1 (x_1+y_2)).\nn
\eea
\subsection{Stable point}
Let us analyze the general rectangular case. We consider the composition of several star-triangle transformation of the form
\bea
\Phi: (t,x,y)\to (t',x',y')\nn
\eea
with
\bea
t'=tx/(t+x+txy).\nn
\eea
The important observation is that the transformation $T_{x,y}:t\to t'$ is a Mobius transformation of hyperbolic type with $t=0$ a stable point . The composition of such transformations
\bea
T_{\overline{x},\overline{y}}=T_{x_n,y_n}\circ\ldots \circ T_{x_1,y_1}\nn
\eea
is of the same type, it is defined by a triangular matrix, hence for generic choice of parameters it has two stable points. One of the stable points is $0$. We are interested in the second stable point 
\bea
t(\{x_1,\ldots,x_n\},\{y_1,\ldots,y_n\}).\nn
\eea
Let us introduce the matrix of the corresponding affine transformation
\bea
A(x,y)=\left(
\begin{array}{cc}
x & 0\\
1+x y & y
\end{array}
\right).
\eea
The stable points of the transformation $T_{\overline{x},\overline{y}}$ are in one-to-one correspondence with the eigenvectors of the matrix given by the product 
\bea
A^{(n)}_{\overline{x},\overline{y}}=\prod_i A(x_i,y_i)=
\left(
\begin{array}{cc}
\prod_i x_i & 0\\
Q^{(n)}(\overline{x},\overline{y}) & \prod_i y_i
\end{array}
\right),
\eea
where
\bea
Q^{(n)}(\overline{x},\overline{y})=\sum_{a=1}^n\left(\prod_{k=1}^{a-1}x_{k} \prod_{k=a+1}^{n} y_{k}\right) 
+x_{1} y_{n}\sum_{a=1}^n\left(\prod_{k=1}^{a-1}x_{k+1} \prod_{k=a+1}^{n} y_{k-1}\right). \nn
\eea
\begin{remark}
In this representation we see that the transformation has two stable points if $\prod_i x_i \ne \prod_i y_i.$
\end{remark}
\begin{lemma}
\bea
\label{t}
t(\overline{x},\overline{y})=\frac {\prod_i x_i-\prod_i y_i}{Q^{(n)}(\overline{x},\overline{y})}.
\eea
\end{lemma}

\subsection{Transformation $r_i$}
Let us calculate the transformations of the variables $X,Y.$  By virtue of the star-triangle transformation we have:
\bea
t_{i+1} &=&  t_i x_i/(t_i+y_i+t_i x_i y_i),\nn\\
x'_i &=& t_i+y_i+t_i x_i y_i,\nn\\
y'_i &=&  x_i y_i/(t_i+y_i+t_i x_i y_i).\nn
\eea
In projective coordinates $u_i/v_i=t_i$ the transformation $t_i \to t_{i+1}$ takes the form
\bea
\left(\begin{array}{c}
u_{i+1}\\
v_{i+1}
\end{array}\right)=A(x_i,y_i)\left(\begin{array}{c}
u_{i}\\
v_{i}
\end{array}\right)=\prod_{k=1}^i A(x_k,y_k)\left(\begin{array}{c}
u_{1}\\
v_{1}
\end{array}\right).
\eea
Hence we obtain the formula
\bea
t_{i+1}=\frac{t \prod_{k=1}^i x_k}{t Q^{(i)}(\overline{x},\overline{y})+\prod_{k=1}^i y_k},\nn
\eea
where $t$ is from equation \ref{t}. Such expressions define the transformation on $X,Y$
\bea
x'_i=x_i t_i/t_{i+1}; \qquad y'_i=y_i t_{i+1}/t_i.\nn
\eea
Let us work out the formula for $t_i$
\bea
\frac 1 {t_{i+1}}=\frac {Q^{(i)}\left(\prod_{k=1}^n x_k-\prod_{k=1}^n y_k\right)+Q^{(n)}\prod_{k=1}^i y_k}{\prod_{k=1}^i x_k \left(\prod_{k=1}^n x_k-\prod_{k=1}^n y_k\right) }=\frac {P^{(i+1)}}{\prod_{k=1}^n x_k-\prod_{k=1}^n y_k}.\nn
\eea
The expression $P^{(i+1)}$ is in fact a polynomial:
\bea
P^{(i+1)}&=&Q^{(i)}\prod_{k=i+1}^n x_k\nn\\
&+&\prod_{k=1}^i y_k \sum_{a=i+1}^n 
\left(\prod_{k=i+1}^{a-1} x_k \prod_{k=a+1}^n y_k + x_{i+1} y_n \prod_{k=i+1}^{a-1} x_{k+1}\prod_{k=a+1}^n y_{k-1} \right)\nn
\eea
which can be simplified as follows:
\bea
P^{(i+1)}=\sum_{a=1}^n\left(\prod_{k=1}^{a-1} x_{i+k}\prod_{k=a+1}^{n}y_{i+k} +x_{i+1} y_i \prod_{k=1}^{a-1} x_{i+k+1}\prod_{k=a+1}^{n}y_{i+k-1}\right)
\eea
We can now define the transformations $r_j$ which act on the $j-th$ and $(j+1)-th$ rows of a rectangular lattice transforming the parameters in these nodes $x_{j,i}$ and $x_{j+1,i}$ by the formulas:
\bea
&r_j(x_{j,i})&=x_{j+1,i} \frac {P_{j,i}}{P_{j,i+1}};\nn\\
&r_j(x_{j+1,i})&=x_{j,i} \frac {P_{j,i+1}}{P_{j,i}};\nn\\
&r_j(x_{k,i})&=x_{k,i} \qquad \mbox{if} \quad k \ne j,j+1;\nn
\eea
where
\bea
\label{pij}
P_{j,i}&=&\sum_{a=1}^n\left(\prod_{k=1}^{a-1}x_{j,i+k} \prod_{k=a+1}^{n} x_{j+1,i+k}\right) \nn\\
&+&x_{j,i+1} x_{j+1,i}\sum_{a=1}^n\left(\prod_{k=1}^{a-1}x_{j,i+k+1} \prod_{k=a+1}^{n} x_{j+1,i+k-1}\right).
\eea
In these formulas we always suppose the cyclic indices $x_{j+m,i}=x_{j,i+n}=x_{j,i}.$
In fact, 
\bea
P_{j,i}=P^{(i+1)}(\overline{x}_j,\overline{x}_{j+1}).\nn
\eea
\begin{remark}
The first summand of \ref{pij} coincides with the formula $(2)$  in \cite{Yam}.
\end{remark}
In a similar way one could define the action on the pairs of columns. Let us denote these transformations by $s_i.$
In terms of the vertex representation for the electric network one obtains quite a simple theorem:
\begin{theorem}
\label{Th_commute}
The transformations $\{r_i\}$ and $\{s_j\}$ define the actions of the symmetric groups $S_m$ and $S_n,$ moreover these actions commute with each other.
\end{theorem}


\begin{remark}
Making the substitution $x_i=\varepsilon \xi_i, y_i=\varepsilon \zeta_i$ and passing to the limit  $\varepsilon\to 0$ we recover the situation studied in \cite{Yam}.
\end{remark}
\begin{remark}
It can be shown as well that
the maps similar to $r_i$ act on the electrical variety corresponding to the standard graph as the figure \ref{mgraphs2} illustrates.
\begin{figure}[h!]
\center
\includegraphics[width=70mm]{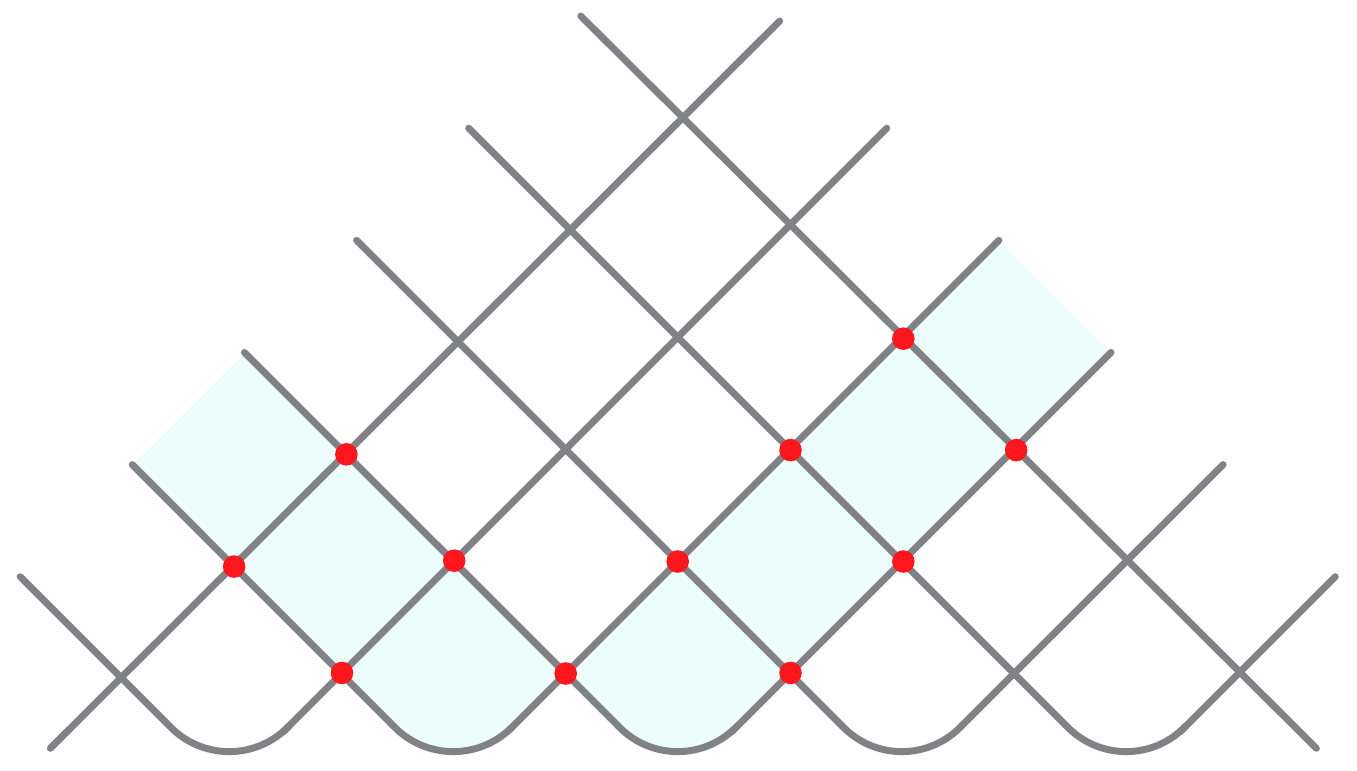}
\caption{Action on standard graphs}
\label{mgraphs2}
\end{figure} 

\end{remark}

\section{Further developments}
We have indicated already a number of ways the results of this paper could be extended. Now we want to suggest a few areas of possible applications of our approach to studying the electrical varieties.

There is a natural connection between electrical networks and so called reversible Markov chains. An example of such a chain is the symmetric graph random walk which, in each step, jumps to a randomly chosen graph 
neighbour at equal probability. This connection is studied in a number of papers and books see \cite{ElecPr}. Interpreting the mathematical structures we have found in this paper in the setup of the Markov chains might be an interesting task.

There is a very intrinsic relationship between the theory of the  electrical networks, the Ising model, the dimer model and the Hopfield neural network model. Just like the random walk model plays the role of the relaxation procedure for the potential distribution problem of the electric network, the dynamics of the Boltzmann machine does the same for the network with not necessarily positive conductivities. We hope that there are similar generalized cluster algebra structures on varieties of neural networks and Ising non homogeneous models.

The theory of electrical networks is a part of discrete harmonic analysis. The methods of the theory of the vertex integrable statistical models we introduced in this paper may therefore find applications in discrete harmonic analysis as well.

The challenging problem is to find the interpretation of the Lusztig-type varieties for the trigonometric solutions for the Zamolodchikov equation from \cite{SS}.

\end{document}